\documentstyle[12pt]{article}
\begin{document}
\def\be{\begin{equation}}
\def\ee{\end{equation}}
\def\ba{\begin{array}}
\def\ea{\end{array}}
\def\bea{\begin{eqnarray}}
\def\eea{\end{eqnarray}}
\parskip=6pt
\baselineskip=20pt
{\raggedleft{{ ASITP}-94-35\\}}
{\raggedleft{{ August,} 1994.\\}}
\bigskip
\bigskip
\bigskip
\medskip
\centerline{\Large \bf A Gauge Field Model On }
\centerline{\Large\bf $SU(2)_L\times SU(2)_R\times U(1)_Y \times
\pi_4(G_{YM}) $
\footnote{ Work supported in part by
The National Natural Science Foundation of China.}}
\vspace{10ex}
\vspace{10ex}
\centerline{\large Bin Chen\footnote{Email: cb@itp.ac.cn}~~ Han-Ying
Guo\footnote{Email: hyguo@itp.ac.cn}~~ Hong-Bo Teng\footnote{Email:
tenghb@itp.ac.cn}~~ and~ Ke Wu\footnote{Email: wuke@itp.ac.cn}}
\vspace{1.5ex}
\centerline{  Institute of Theoretical Physics, Academia Sinica,
P. O. Box 2735, Beijing 100080, China.}
\vspace{8ex}
\vspace{8ex}

 \centerline{Abstract}
{\it We reconstruct the Lagrangian of a
left-right symmetric model with the gauge group $SU(2)_L\times SU(2)_R\times
U(1)_Y \times
\pi_4(SU(2)_L\times SU(2)_R\times U(1)_Y) $. The Higgs fields appear as gauge
fields on discrete gauge group $\pi_4(SU(2)_L\times SU(2)_R\times U(1)_Y) $ and
are assigned in a way complying with the principle that both the
original gauge group $G_{YM}$ and the discrete group $\pi_4(G_{YM})$ should be
taken as gauge groups in sense of non-commutative geometry.
}

\newpage

\section{Introduction}

Very recently, an $SU(2)$ generalized gauge field model
has been constructed\cite{new}. In this model, the Yang-Mills
gauge group
$SU(2)$ and its
fourth homotopy group $\pi_{4}(SU(2)) = Z_{2}$ are dealt with on the equal
footing in the sense of non-commutative differential geometry\cite{connes1}. It
is
remarkable that not only the Higgs mechanism is automatically included in the
model but also it survives quantum correlations since the spontaneous symmetry
breaking breaks down both $SU(2)$ and $\pi_{4}(SU(2))$. The later is different
from Connes' NCDG approach to the particle model building\cite{connes1}.
In \cite{new1}, this model is generalized to the Weinberg-Salam model and the
standard model.
The reason why the fourth homotopy group plays this important role lies in the
fact that in these cases the base manifold is the $4-$dimensional spacetime
which may be compactified to $S^4$ and there is a kind of non-trivial
$SU(2)_L$ gauge
transformations
which are topologically inequivalent to the identity. That means
there does exist an internal gauge symmetry which we used to neglect by
considering the infinitesimal transformation of a Lie gauge group. Taking
into account this fact, both $G_{YM}$ and $\pi_4(G_{YM})$ should be taken
as gauge
groups on the equal footing, where $G_{YM}$ is the original gauge group and
$\pi_4(G_{YM})$ is the fourth homotopy group of $G_{YM}$. Then using the
mathematic structure in\cite{connes1}, \cite{sitarz}, the gauge fields will
have two parts:
the gauge fields on $G_{YM}$ which are the same as before and the gauge fields
on discrete group $\pi_4(G_{YM})$ which appear as the Higgs fields. And the
Yukawa coupling and Higgs mechanism will appear as a natural result.

 It should be mentioned that since the
discrete symmetry is spontaneously broken down at the same time with the
continuous gauge symmetry, there is no need to concern about this discrete
symmetry
when we quantize the theory. On the other hand, other
approaches\cite{hhjk1,hhjk2,cb,ali,co}
does not survive the quantum correlation.

If this principle is true, it should be
applicable to all $4-D$ gauge theory models with non-trivial fourth homotopy
groups of the gauge groups. One of them is the left-right symmetric model.
 As is well known, a missing link in the standard model is that the
$V-A$ structure of currents is put
by hand. In the middle 70's,  Pati, Salam and Mohapatra\cite{pasa,mopa}
proposed a gauge theory model based on $SU(2)_L\times SU(2)_R\times U(1)_Y$
which is totally
left-right symmetric before symmetry breaking. They showed the right-handed
charged gauge meson $W_R^+$ can be made much heavier than the left-handed
$W_L^+$ and the $V-A$ structure of weak interaction can be
regarded as a low energy phenomenon which should disappear at $10^3 Gev$
or higher. And in the limit of infinitely heavy $W_R^+$, the predictions
are the same as the standard model, as far as the charged and neutral currents
interactions are concerned. In other words, such theories are
indistinguishable from the standard model at low energies.

 In this paper, we study such a model of left-right
symmetry based on
$SU(2)_L\times SU(2)_R\times U(1)_Y \times \pi_4(SU(2)_L\times
SU(2)_R\times U(1)_Y) $. We choose and assign the Higgs fields according to the
principle mentioned above. We show it is the minimum Higgs assignment
 which is necessary to break the gauge group down to $U(1)_{em}$ and the three
Higgs fields included can be regarded
as gauge fields along the three directions of the tangent space of
$\pi_4(SU(2)_L\times SU(2)_R\times U(1)_Y) = Z_2\oplus Z_2$.

 We first show the differential calculus on $Z_2\oplus Z_2$ in section $2$.
Then  in section $3$ we construct the
Lagrangian of this model. Finally we end
 with conclusions.

\section{Differential Calculus on $Z_2\oplus Z_2$}

We will not present the general structure of non-commutative
geometry\cite{sitarz} in this section. But since $\pi_4(SU(2)_L\times
SU(2)_R\times U(1)_Y) = Z_2\oplus Z_2$, we will briefly introduce some
notations
and relations of the differential calculus on $Z_2\oplus Z_2$. Let's write the
four elements of $Z_2\oplus Z_2$ as \[ (e_1, e_2),~ (r_1, e_2),~ (e_1, r_2),
{}~ (r_1, r_2).\] And the group multiplication is
\be
(g_1, g_2)(h_1, h_2)=(g_1h_1, g_2h_2).
\ee
Let $\cal A$ be the algebra of complex valued functions on $Z_2\oplus Z_2$.
 The derivative on $\cal A$ is defined as
\be
\partial_gf=f-R_{g}f   \hspace{4ex} g\in Z_2\oplus Z_2 ,\hspace{1ex} f\in\cal A
\ee
with  $R_{g}f(h)=f(hg)$. We will write $\partial_i$ and $R_i$ for convenience
where $i=1, 2, 3$ refers to $(r_1, e_2), (e_1, r_2), (r_1, r_2)$ respectively.

The bases of space of one forms are $\chi^1, \chi^2, \chi^3$ which are defined
with
\be
\chi^i(\partial_j)=\delta^i_j   \hspace{4ex} i, j=1,2,3.
\ee
One can easily find that the following relations hold
\bea \label{alrelation}
\partial_1\partial_2&=&\partial_1+\partial_2-\partial_3 \nonumber\\
\partial_1\partial_1&=&2\partial_1 \nonumber\\
\partial_1\partial_2&=&\partial_2\partial_1 \\
d_{Z_2\oplus Z_2}\chi^1&=&-\chi^1\otimes\chi^2-\chi^1\otimes\chi^3+\chi^2
\otimes\chi^3-2\chi^1\otimes\chi^1-\chi^2\otimes\chi^1-\chi^3\otimes
\chi^1+\chi^3\otimes\chi^2 \nonumber \\
\vdots &&~~ \mbox{and similar eqs under permutations (1,2,3) and (2,1,3)}.
\nonumber
\eea
 In order to get a left-right symmetric Lagrangian before symmetry breaking,
we let the metric to be symmetric under $1\leftrightarrow2$
\be \label{metric1}
\ba{l}
<\chi^1, \chi^1>=<\chi^2, \chi^2>=\eta ~~; \hspace{2ex} <\chi^3, \chi^3>=
{\eta}^{\prime} \\
<\chi^i, \chi^j>=0, \hspace{4ex} i\neq j.
\ea
\ee
And let
\be \label{metric2}
<\chi^i\otimes\chi^j, \chi^k\otimes\chi^l>=a<\chi^j, \chi^k><\chi^i,
\chi^l>+b<\chi^i, \chi^k><\chi^j, \chi^l>.
\ee
with $a$,$b$ two constants.

\section{Gauge Theory}

Since there are three bases of the space of one forms, we take in three Higgs
fields in this model
\be \label{higgsinclude}
\Phi=\left(\ba{cc} \phi^0_1&\phi^+_1\\ \phi^-_2&\phi^0_2 \ea \right);
\Delta_L=\left(\ba{c}\chi_L^+\\ \chi_L^0\ea\right); \Delta_R=\left(\ba{c}
\chi_R^+\\ \chi_R^0\ea\right)\ee
 which belong to $(\frac{1}{2}, {\frac{1}{2}}^*, 0), (\frac{1}{2}, 0, 1)$
and $(0, \frac{1}{2}, 1)$ respectively. And this is also the minimum choice to
give
 necessary symmetry breaking. We also choose the Lagrangian to be invariant
under $L\leftrightarrow R$ transformation so we let $g_L=g_R=g$.

All fields are regarded as elements of function space on $SU(2)_L\times
SU(2)_R\times U(1)_Y \times \pi_4(SU(2)_L\times SU(2)_R\times U(1)_Y)$.
 We reasonably postulate that an $SU(2)$ singlet is invariant under
$\pi_4(SU(2))$, that means $R\stackrel{R_1}{\rightarrow}R$,
$L\stackrel{R_2}{\rightarrow} L$, etc. Here $L(R)$ is left(right)-handed
fermion
doublet. Hence we write, for fermions
\be \label{fermion}
\ba{ll}
L(x,e_1,e_2)=L;&L(x,r_1,e_2)=R_1L\equiv L^{r^{1}}\\
L(x,e_1,r_2)=R_2L=L;&L(x,r_1,r_2)=R_3L=L^{r^1}  \\
R(x,e_1,e_2)=R;&R(x,r_1,e_2)=R_1R=R\\
R(x,e_1,r_2)=R_2R\equiv R^{r^{2}};&R(x,r_1,r_2)=R_3R=R^{r^2}
\ea
\ee
where $L=\left(\ba{c} {\nu_l} \\ l \ea\right)_L$ or $\left(\ba{c} {u^i}
\\ {d^i} \ea\right)_L$, which belong to $(\frac{1}{2},0,-1)$ and
$(\frac{1}{2},0,\frac{1}{3})$ respectively, and $R=\left(\ba{c} {\nu_l} \\ l
\ea\right)_R$ or $\left(\ba{c} {u^i}\\ {d^i} \ea\right)_R$, belong to
$(0,\frac{1}{2},-1)$ and $(0,\frac{1}{2},\frac{1}{3})$ respectively.\\ \\
For gauge fields
\be \label{gaugefield}
\ba{ll}
L_{\mu}(x,e_1,e_2)=L_{\mu};&L_{\mu}(x,r_1,e_2)=R_1L_{\mu}\equiv
L_{\mu}^{r^{1}}\\
L_{\mu}(x,e_1,r_2)=R_2L_{\mu}=L_{\mu};&L_{\mu}(x,r_1,r_2)=R_3L_{\mu}
=L_{\mu}^{r^1} \\
R_{\mu}(x,e_1,e_2)=R_{\mu};&R_{\mu}(x,r_1,e_2)=R_1R_{\mu}=R_{\mu}\\
R_{\mu}(x,e_1,r_2)=R_2R_{\mu}\equiv R_{\mu}^{r^{2}};&R_{\mu}(x,r_1,r_2)
=R_3R_{\mu}=R_{\mu}^{r^2}
\ea
\ee
where $L_\mu=-ig\frac{\tau^i}{2}W_\mu^{Li}-ig^{\prime}\frac{Y}{2}B_\mu$,
$R_\mu=-ig\frac{\tau^i}{2}W_\mu^{Ri}-ig^{\prime}\frac{Y}{2}B_\mu$.\\
For Higgs fields
\be \label{higgs}
\ba{ll}
\Phi(x,e_1,e_2)=\Phi;&\Phi(x,r_1,e_2)=R_1\Phi\equiv\Phi^{r^1};\\
\Phi(x,e_1,r_2)=R_2\Phi\equiv\Phi^{r^2};&\Phi(x,r_1,r_2)=R_3\Phi
\equiv\Phi^{r^3}\\
&\\
\Delta_L(x,e_1,e_2)=\Delta_L;&\Delta_L(x,r_1,e_2)=R_1\Delta_L
\equiv\Delta_L^{r^1};\\
\Delta_L(x,e_1,r_2)=R_2\Delta_L=\Delta_L;&\Delta_L(x,r_1,r_2)
=R_3\Delta_L=\Delta_L^{r^1}\\
&\\
\Delta_R(x,e_1,e_2)=\Delta_R;&\Delta_R(x,r_1,e_2)=R_1\Delta_R
=\Delta_R;\\
\Delta_R(x,e_1,r_2)=R_2\Delta_R\equiv\Delta_R^{r^2};&\Delta_R(x,r_1,r_2)
=R_3\Delta_R=\Delta_R^{r^2}
\ea
\ee
\\
And we will not need the detail information about how the
fields transform.

Now taking into account the representations which the fields belong to, we
assign these fields into three sectors. On point $(e_1,e_2)$, we have
\be
\label{assign}
\ba{l}
\Psi(x,e_1,e_2)=\left(\ba{c} L\\R\\0\ea \right)=\Psi(x);~~A_\mu(x,e_1,e_2)
=\left(\ba{ccc} L_\mu&&\\&R_\mu&\\&&0 \ea\right)=A_\mu(x)\\ \\
\phi_1=\left(\ba{ccc} \alpha_1&&-\Delta_L\\&\alpha_1&\\-
\Delta_L^{{\dag}r_1}&&\alpha_1 \ea\right);~~\phi_2=
\left(\ba{ccc} \alpha_2&&\\&\alpha_2&-\Delta_R\\
&-\Delta_R^{{\dag}r_2}&\alpha_2 \ea \right);~~\phi_3
=\left(\ba{ccc} \alpha_3&-\Phi^{r_2}&\\-\Phi^{{\dag}r_1}&\alpha_3&\\&&\alpha_3
\ea \right)
\ea
\ee

While on other points of $Z_2\oplus Z_2$ these fields can be easily written out
according to (\ref{fermion}),(\ref{gaugefield}),(\ref{higgs}).
Here we see the generalized gauge fields including both the original gauge
fields and the Higgs fields are arranged into $5\times 5$ matrices which belong
to
the representation $[(\frac{1}{2},0)\oplus (0,\frac{1}{2})\oplus (0,0)]\otimes
[(\frac{1}{2}^*,0)\oplus (0,\frac{1}{2}^*)\oplus (0,0)]$ of $SU(2)_L\times
SU(2)_R$.

It should be mentioned that assignment (\ref{assign}) not only assigns
the fields to the points of $Z_2\oplus Z_2$ but also gives certain
matrices arrangement.
 Such an arrangement can only be viewed as a working hypothesis at this moment
and
sometimes one should avoid
 certain extra constraints coming from this arrangement.

It is easy to see
\be
\phi_i^{\dag} =\phi_i^{r_i} \hspace{3ex} i=1, 2, 3.
\ee
The connection one-form on $SU(2)_L\times
SU(2)_R\times U(1)_Y \times \pi_4(SU(2)_L\times SU(2)_R\times U(1)_Y) $ is
\be
A(x,h)=A_\mu(x,h)dx^\mu+\sum_{i=1}^{3}\frac{1}{\alpha_i}\phi_i(x,h)\chi^i.
\ee
We see that $\phi_1$,$\phi_2$,$\phi_3$ are Higgs fields assigned to three
directions of $\Omega^1$, the space of one-forms on the discrete group. Such an
assignment is quite a natural choice as far as the principle of generalized
gauge group is concerned.

The generalized curvature two-form reads
\be
F(h)
\ba[t]{l} =dA+A\otimes A\\
=\frac{1}{2}F_{\mu\nu}dx_\mu\wedge dx_\nu +\sum_{i=1}^{3}
\frac{1}{\alpha_i}F_{r_i\mu}\chi^idx^\mu +\sum_{i,j=1}^{3}
\frac{1}{\alpha_i}\frac{1}{\alpha_j}F_{r_ir_j}\chi^i\chi^j
\ea\ee
here $d=d_M+d_{Z_2\oplus Z_2}$. After some calculation, we get
\be
\ba{l}
F_{\mu\nu}=\left(\ba{ccc} L_{\mu\nu}&&\\&R_{\mu\nu}&\\&&0 \ea\right)\\ \\
F_{r_1\mu}=D_\mu\Phi_1=\left(\ba{ccc} &&D_\mu\Delta_L\\&0&\\
({D_\mu\Delta_L})^{{\dag}r_1}&& \ea\right)\\ \\
F_{r_2\mu}=D_\mu\Phi_2=\left(\ba{ccc} 0&&\\&&D_\mu\Delta_R\\
&({D_\mu\Delta_R})^{{\dag}r_2}& \ea\right)\\ \\
F_{r_3\mu}=D_\mu\Phi_3=\left(\ba{ccc} &{D_\mu\Phi}^{r_2}&\\(
{D_\mu\Phi})^{{\dag}r_1}&&\\&&0 \ea\right)
\ea
\ee
and
\be
\ba{l}
F_{r_1r_1}=\Phi_1\Phi_1^{\dag}-\alpha_1^2\\ \\
F_{r_1r_2}=\Phi_1\Phi_2^{r_1}-\frac{\alpha_1\alpha_2}{\alpha_3}\Phi_3\\
\vdots \hspace{3ex} \mbox{and similar eqs under permutations
(1,2),(1,3),(2,3),(1,2,3),(1,3,2).}
\ea
\ee
\\
where $\Phi_i=\alpha_i-\phi_i$, $L_{\mu\nu}=-ig\frac{\tau^i}{2}
W_{\mu\nu}^{Li}-ig^{\prime}\frac{Y}{2}B_{\mu\nu}$, $R_{\mu\nu}=
-ig\frac{\tau^i}{2}W_{\mu\nu}^{Ri}-ig^{\prime}\frac{Y}{2}B_{\mu\nu}$, and
\be
\ba{l}
D_\mu\Delta_L=\partial_\mu\Delta_L+L_\mu\Delta_L\\
D_\mu\Delta_R=\partial_\mu\Delta_R+R_\mu\Delta_R\\
D_\mu\Phi =\partial_\mu\Phi +L_\mu\Phi -\Phi R_\mu .
\ea
\ee

Using (\ref{metric1}), (\ref{metric2}) and
\be
{\cal L}=<F(h), {\bar {F}}(h)>
\ee
we have the bosonic sector of the Lagrangian
\be
{\cal L}_{YM-H}(x)=
\ba[t]{l} -\frac{1}{4N_L}TrL_{\mu\nu}L_{\mu\nu}-\frac{1}{4N_R}
TrR_{\mu\nu}R_{\mu\nu}\\ \\
	-\frac{2}{N}\frac{\eta}{\alpha^2}
[Tr(D_\mu\Delta_L)(D_\mu\Delta_L)^{\dag}+
Tr(D_\mu\Delta_R)(D_\mu\Delta_R)^{\dag}]-
\frac{2}{N^{\prime}}\frac{\eta^{\prime}}{\alpha^{{\prime}2}}
Tr(D_\mu\Phi)(D_\mu\Phi)^{\dag}\\ \\
	-\frac{2}{N}(\frac{\eta}{\alpha^2})^2
[Tr(\Delta_L\Delta_L^{\dag}-\alpha^2)^2+
Tr(\Delta_R\Delta_R^{\dag}-\alpha^2)^2]
-\frac{2}{N^{\prime}}(\frac{\eta^{\prime}}{\alpha^{{\prime}2}})^2
Tr(\Phi\Phi^{\dag}-\alpha^{{\prime}2})^2\\ \\
	-\frac{1}{N_1}\frac{\eta}{\alpha^2}\frac{\eta}{\alpha^2}
\{2a\Delta_L^{\dag}\Delta_L\Delta_R^{\dag}\Delta_R-
2(a+b)\frac{\alpha^2}{\alpha^{\prime}}
(\Delta_L^{\dag}\Phi\Delta_R+\Delta_R^{\dag}\Phi^{\dag}\Delta_L)+
4(a+b)(\frac{\alpha^2}{\alpha^{\prime}})^2Tr(\Phi\Phi^{\dag})\}\\ \\
	-\frac{1}{N_1^{\prime}}\frac{\eta}{\alpha^2}
\frac{\eta^{\prime}}{\alpha^{{\prime}2}}
\{2a\Delta_L^{\dag}\Phi\Phi^{\dag}\Delta_L-
2(a+b)\alpha^{\prime}(\Delta_L^{\dag}\Phi\Delta_R+
\Delta_R^{\dag}\Phi^{\dag}\Delta_L)+
4(a+b)\alpha^{{\prime}2}\Delta_R^{\dag}\Delta_R\}\\ \\
	-\frac{1}{N_1^{\prime}}\frac{\eta}{\alpha^2}
\frac{\eta^{\prime}}{\alpha^{{\prime}2}}
\{2a\Delta_R^{\dag}\Phi^{\dag}\Phi\Delta_R-
2(a+b)\alpha^{\prime}(\Delta_L^{\dag}\Phi\Delta_R+
\Delta_R^{\dag}\Phi^{\dag}\Delta_L)+
4(a+b)\alpha^{{\prime}2}\Delta_L^{\dag}\Delta_L\}\\ \\
	+const.
\ea
\ee

Here we have set $\alpha_1=\alpha_2=\alpha$, $\alpha_3=\alpha^{\prime}$
 in order to get a left-right symmetric Lagrangian and inserted
some normalization constants $N_L, N_R, N, N^{\prime},
N_1, N_1^{\prime}$ to avoid extra constraints coming from our
 arrangement of the fields. Since ${\cal L}$ is invariant under $SU(2)_L\times
SU(2)_R\times U(1)$ transformation
 and so that is independent of the elements of $Z_2\oplus Z_2$,
there is no need to take the Haar integral.

As to the fermionic sector of the Lagrangian, we have
\be
{\cal L}_F(x)=-{\bar L}\gamma_\mu(\partial_\mu+L_\mu)L-{\bar R}
\gamma_\mu(\partial_\mu+R_\mu)R-\lambda({\bar L}\Phi R+{\bar R}\Phi^{\dag}L)
\ee
where $\lambda$ is the Yukawa coupling constant. We see that only $\Phi$
 has contribution to fermions masses, since $\Delta_L, \Delta_R$ can not give a
group singlet in Yukawa coupling.

After choosing proper normalization constants, we rewrite the above Lagrangian
as
\be
{\cal L}_{YM-H}(x)=
\ba[t]{l} -\frac{1}{4}W_{\mu\nu}^{Li}W_{\mu\nu}^{Li}-
	\frac{1}{4}W_{\mu\nu}^{Ri}W_{\mu\nu}^{Ri}-
	\frac{1}{4}B_{\mu\nu}B_{\mu\nu}\\ \\
	-[Tr(D_\mu\Delta_L)(D_\mu\Delta_L)^{\dag}+
	Tr(D_\mu\Delta_R)(D_\mu\Delta_R)^{\dag}]-
	Tr(D_\mu\Phi)(D_\mu\Phi)^{\dag}\\ \\
	-V(\Phi,\Delta_L, \Delta_R)
\ea
\ee
and
\be
V(\Phi,\Delta_L,\Delta_R)=
\ba[t]{l} \rho_1(\Delta_L^{\dag}\Delta_L\Delta_L^{\dag}\Delta_L+
\Delta_R^{\dag}\Delta_R\Delta_R^{\dag}\Delta_R)- \mu_1^2(\Delta_L^{\dag}
\Delta_L+\Delta_R^{\dag}\Delta_R)\\ \\
	+\rho_2Tr(\Phi\Phi^{\dag}\Phi\Phi^{\dag})
-\mu_2^2Tr(\Phi\Phi^{\dag})\\ \\
	+\rho_3\Delta_L^{\dag}\Delta_L\Delta_R^{\dag}\Delta_R
+\rho_4Tr(\Delta_L^{\dag}\Phi\Phi^{\dag}\Delta_L+
\Delta_R^{\dag}\Phi\Phi^{\dag}\Delta_R)\\ \\
	-\mu_3^2(\Delta_L^{\dag}\Phi\Delta_R+\Delta_R^{\dag}\Phi\Delta_L)

\ea
\ee

The breaking pattern of such a Higgs potential is well-known. We refer the
readers to \cite{senjan} by G. Senjanovic for a detail discussion.

\section{Conclusions}

Based on a generalized
gauge principle, we constructed an $SU(2)_L\times SU(2)_R\times U(1)_Y$
model with $\pi_4(SU(2)_L\times SU(2)_R\times U(1)_Y)$ taken as discrete gauge
symmetry. The Higgs mechanism is automatically included in this generalized
gauge theory model. And the choice and assignment of Higgs fields seems quite
 natural when the principle of generalized gauge group is taken into account.
Besides, since we take the homotopy group as generalized gauge
group, the discrete symmetry is broken synchronously with the
continuous symmetry. So we get the same version as an ordinary Yang-Mills
model. That means we do not need to concern about this discrete symmetry when
we quantize the model.

\end{document}